\documentclass[twocolumn,showpacs,amsmath,prl]{revtex4-1}

\usepackage{graphicx,bm}
\usepackage{epsfig,psfrag}
\usepackage{amsmath}
\usepackage{amssymb}
\usepackage{amsbsy}
\usepackage{amsthm}
\usepackage{amsfonts}
\usepackage{wasysym}
\usepackage{bbm}
\usepackage{tabularx}
\usepackage{euscript}
\usepackage{color}
\usepackage{enumerate}
\usepackage{amsfonts}
\usepackage{exscale}
\usepackage{bbold}
\usepackage{float}

\usepackage[colorlinks,citecolor=blue]{hyperref}

\newcommand{\Ref}[1]{Ref.~\cite{#1}}
\newcommand{\Eq}[1]{equation~(\ref{#1})}
\newcommand{\Fig}[1]{Fig.~\ref{#1}}

\newcommand{\ra}{\rightarrow}
\newcommand{\s}{{\sigma}}
\def\bea{\begin{eqnarray}}
\def\eea{\end{eqnarray}}
\def\bra#1{\left\langle#1\right|}
\def\ket#1{\left|#1\right\rangle}
\def\avg#1{\left\langle#1\right\rangle}

\def\Eq#1{Eq.~(\ref{#1})}
\def\Fig#1{Fig.~\ref{#1}}

\begin{document}

\title{The nature of effective interaction in cuprate superconductors:
a sign-problem-free
quantum Monte-Carlo study}

\author{Zi-Xiang Li$^{1}$, Fa Wang$^{2,3}$, Hong Yao$^{1,3}$ \& Dung-Hai Lee$^{4,5}$}

\affiliation{
$^1$ Institute for Advanced Study, Tsinghua University, Beijing 100084, China.\\
$^2$ International Center for Quantum Materials, School of Physics, Peking University, Beijing 100871, China.\\
$^3$ Collaborative Innovation Center of Quantum Matter, Beijing, China.\\
$^4$ Department of Physics, University of California, Berkeley, CA 94720, USA.\\
$^5$ Materials Sciences Division, Lawrence Berkeley National Laboratory, Berkeley, CA 94720, USA.}

\begin{abstract}
\end{abstract}

\maketitle

{\bf
Superconductivity is an emergent phenomena in the sense that the energy scale associated with Cooper pairing is generically much lower than the typical kinetic energy of electrons.
Addressing the mechanism of Cooper pairing amounts to determine the effective interaction that operates at low energies. Deriving such an interaction from a bottom-up approach has not been possible for any superconductor, especially strongly correlated ones.
Top-down approaches, where one assumes an effective interaction, is plagued with the difficulty of extracting the implied electronic instabilities without uncontrolled approximations. These facts severely hinder our ability to determine the pairing mechanism for high temperature superconductors. Here we perform large-scale sign-problem-free quantum Monte-Carlo simulations on an effective theory, featured with antiferromagnetic and nematic fluctuations, to study the intertwined antiferromagnetic, superconducting, and charge density wave instabilities of the cuprates. Our results suggest the inclusion of nematic fluctuations is essential in order to produce the observed type of charge density wave ordering. Interestingly
we find that the d-wave Cooper pairing is enhanced by nematic fluctuations.}

In the last few years it is established that in addition to the antiferromagnetic (AF) and superconducting (SC) orders both electron and hole doped cuprates superconductors exhibit the propensity toward charge density wave (CDW) order\cite{Tranquada-1995, Wu-2011, Mesaros-2011, Hayden, Keimer, Davis, Comin1, Tabis-2014, Hashimoto, Comin, silva, Gerber-2015} .  These instabilities together with nematicity\cite{Ando,Hinkov,Lawler,Daou} form the so-called ``intertwined orders'' of the cuprate superconductors\cite{Kivelson-2013,Dunghai-2013,Kivelson-2014}.
It is highly demanded to find the correct effective interaction for the cuprates that drives all the above instabilities.

The theoretical progress in the cuprate high temperature superconductors has been hindered by the strong electron-electron correlations. In particular {\it un-biased} calculations with no uncontrolled approximation are extremely rare. Under such circumstance phenomenological approaches based on various degree of approximations and the assumption that antiferromagnetic fluctuation  is mainly responsible for high temperature superconductivity and all its ``intertwined'' orders have been useful for the understanding of the plethora of electronic instabilities in cuprates\cite{Efetov,Kivelson-2014,Kivelson-2015,Dunghai-2013,Laplaca,Subir-2014,Subir-20142,Chubukov-2014}.
\begin{figure}
\centering
\includegraphics[width=8.5cm]{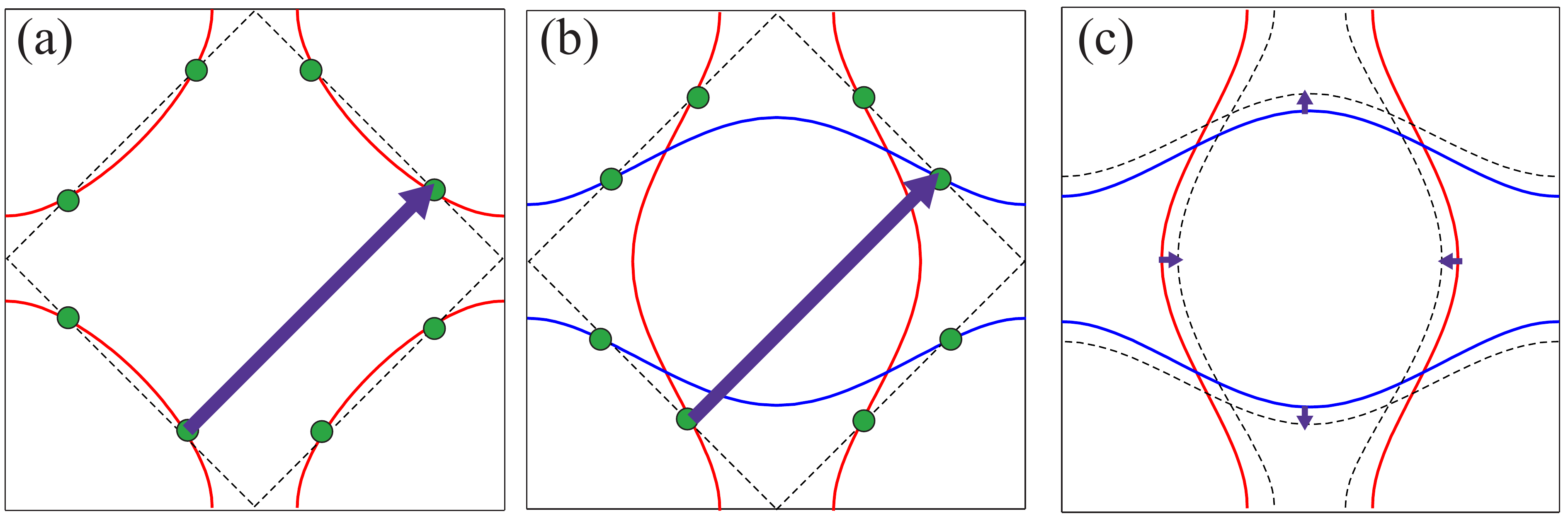}
\caption{(a) A prototypical Fermi surface of the cuprates. The hot spots are denoted by green dots and the purple arrow indicate the AF ordering wavevector. (b) The Fermi surface of the two-band model where the positions of hot spots and the Fermi velocity at hot spots are made to mimic those in the single-band model used to describe the cuprates. The red/blue Fermi surfaces are derived from the bands formed by the $x$ and $y$ orbitals described by the action in \Eq{action}. (c) The nematically distorted Fermi surface.  The arrows indicate the distortion.}
\label{fig1}
\end{figure}

\begin{figure*}[t]
\includegraphics[width=18.0cm]{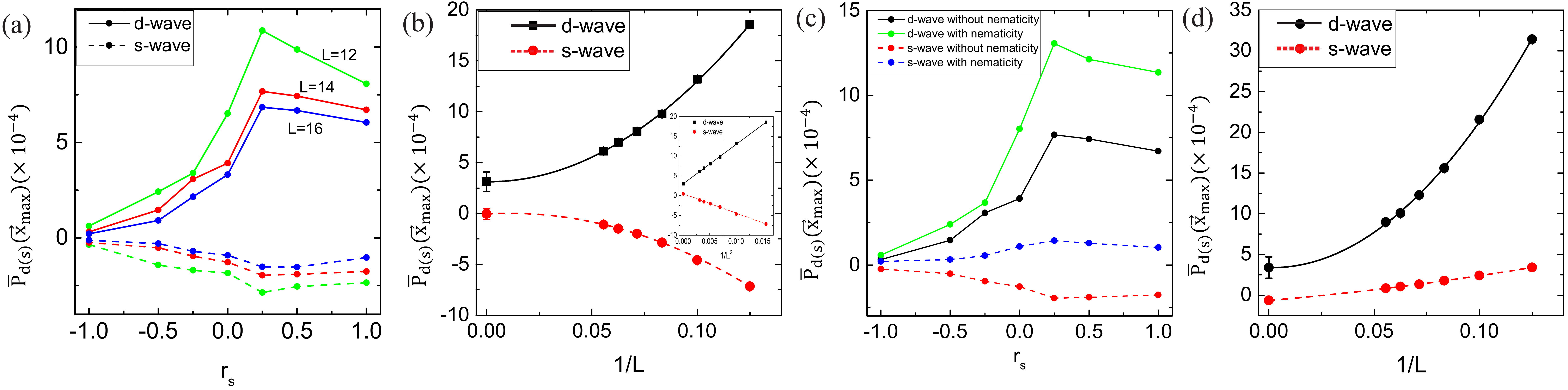}
\caption{
(a) The $s$- and $d$-wave SC pair correlations $\overline P_{d(s)}$ evaluated at maximum separation $\vec x_\textrm{max}=(L/2,L/2)$ for $L=12,14,16$ and various values of $r_s$. The peak of the enhancement for $d$-wave pairing occurs at $r_s\approx 0.25$ which is close to the AF quantum critical point. (b) Both $s$- and $d$-wave pair correlations evaluated at $\vec x_\textrm{max}$ are plotted versus $1/L$ for $L=8,10,12,14,16,18$. By fitting them using $f(1/L) = a + b/L + c/L^2$ and extrapolating to the thermodynamic limit ($L=\infty$), we find that the $d$-wave pairing correlations are extrapolated to a non-zero value, $(3.1\pm 0.8)\times 10^{-4}$, while the $s$-wave pairing correlations to zero within the error bar. Interestingly the finite size dependence of both correlation functions are dominated by $1/L^2$ as shown by the nearly linear dependence of $\overline{P}_{d(s)}$ on $1/L^2$  in the inset of \Fig{fig2}(b). These results clearly indicate that the ground state possesses $d$-wave superconducting long-range order.
(c) The comparison of the $d$- and $s$-wave SC pairing correlations at $\vec x_\textrm{max}$ in the system with $L=14$, without and with nematic fluctuation. In the latter case, we set $r_n=0.5$ which places the system on the disordered side of the nematic transition. The results show an considerably enhanced $d$-wave SC correlation by the nematic fluctuations. (d)  $\overline P_{d(s)}(\vec x_\textrm{max})$ as a function of $1/L$ for $1/L$ for $L=8,\cdots,18$. Here $(r_s,r_n)$ are set to $(0.5,0.5)$. The solid and dashed lines are the best fit using $f(1/L) = a + b/L + c/L^2$. From comparing the results in this panel with that in panel (b), we see the $d$-wave pairing is moderately enhanced by nematic fluctuations. }
\label{fig2}
\end{figure*}

Low energy effective theories involving the AF fluctuations and the electrons around hot spots ({i.e.} the momentum space points situated at the intersections of the normal state Fermi surface and the AF Brillouin zone boundary as shown in \Fig{fig1}(a))
have been used to address the AF quantum phase transition\cite{Abanov-2000,Abanov-2004,Metlitski-2010,Hartnoll-2011}.
Recently an insightful work of Berg {\it et. al.} \cite{berg-2012} has achieved in performing a sign-problem-free finite temperature quantum Monte-Carlo simulation on an effective model featuring the AF fluctuations  and electrons whose band structure shares the same hot spots as the cuprates.  The results suggest that quantum critical AF fluctuation enhances Cooper pair correlations in the $d$-wave channel (although the $d$-wave long-range order has not be numerically observed). Since recent experimental progresses show an incommensurate CDW instability is ubiquitous among all cuprates,it is natural to ask whether such an effective theory can account for the CDW instability as well.

Here we perform large-scale zero-temperature sign-problem-free projector QMC simulations\cite{Sorella-1989, White-1989,Assaad-2005} to answer this question (see supplemtnary information VI for a detailed description). The main results are summarized as follows. (1) We show that the ground state of the effective theory in \Ref{berg-2012} has $d$-wave superconducting {\it long-range order} and the $d$-wave pairing is strongest close to the AF quantum critical point. (2) However, the CDW instability it predicts has the ordering wave vectors {\it inconsistent} with the ones observed experimentally. (3) After introducing  the nematic fluctuation to the effective theory the correct type of CDW instability emerges. (4) In the presence of nematic fluctuations our theory favors uni- rather than bi-directional CDW. (5) The nematic fluctuation  enhances the $d$-wave Cooper pairing\cite{Kivelson-2015}. (6) Robust superconductivity with predominant d-wave pairing symmetry exists in the nematic ordered phase. These results, plus the fact that with nematic fluctuation our theory can naturally account for the observed nematic instability, lead us to conclude that nematic fluctuation is indispensable in ``hot spot theories'' of the cuprates.

A prototypical Fermi surface of cuprates is shown in \Fig{fig1}(a). It originates from a band derived from the copper $3d_{x^2-y^2}$ and the oxygen $2p_{x/y}$ orbitals. The hot spots (the green dots) are the intersections of this Fermi surface and the AF Brillouin zone boundary (the dashed lines). In the same figure the $(\pi,\pi)$ AF ordering wavevector is shown as the purple arrow. The fact that the AF fluctuation scatters electron {\it within the same band} hinders sign-problem-free QMC simulations so far. Following \Ref{berg-2012} we consider a two-band model. The two bands are derived from two orbitals on each site of the square lattice. The Fermi surface (marked red and blue in \Fig{fig1}(b)) of this model features the same hot spots. Moreover up to linear order in the momentum deviation from the hot spots the electron dispersion is  very similar to that of the cuprates. These facts motivate one to think that an effective theory based on this new band structures and the AF fluctuation can be used to simulate the low-energy physics of cuprates when the system is not too far from the AF quantum critical point\cite{berg-2012}.

The AF effective action based on the two-band model consists of the band electrons coupled to an fluctuating AF order parameter by the Yukawa coupling. The details are given in supplementary information I. The most important parameter in this action is $r_s$ which tunes the system across the AF quantum phase transition. More specifically the disordered phase lies in the range $r_s>r_{s,c}$ and the ordered phase requires $r_s<r_{s,c}$, where $r_{s,c}$ marks the AF quantum critical.
Our large-scale projective QMC simulation is carried out on a square lattice with $N=L\times L$ sites. Unless otherwise mentioned we use periodic boundary condition. From the finite-size scaling of the Binder-ratio\cite{binder} associated with the AF order parameters (see supplemental materials), we determine  $r_{s,c}\approx 0.25$ which is consistent with the value obtained in \Ref{berg-2012}. In the following we present the simulation results on the SC and CDW instabilities and their dependence on the nematic fluctuations.\\

\begin{figure*}[t]						
\centering
\includegraphics[width=17.8cm]{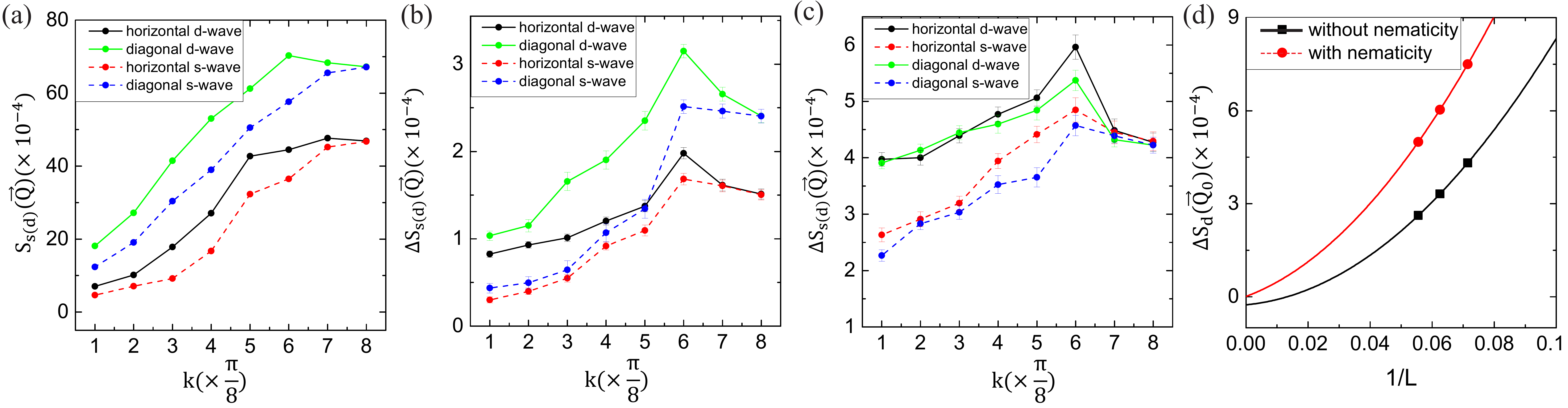}
\caption{
(a) The $s$- and $d$-wave bond CDW structure factor $S_{s(d)}(\vec{Q})$ versus $\vec{Q}=(k,0)$ (horizontal CDW) or $\vec{Q}=(k,k)$ (diagonal CDW) for the system with $L=16$. Here $r_s$ is set to $0.5$. (b)  $\Delta S_{s/d}(\vec Q)$, the difference between  $S_{s/d}(\vec{Q})$ with ($r_s=0.5$) and without the AF fluctuations, versus $\vec{Q}$ for L=16.  (c)  $\Delta S_{s/d}(\vec Q)$, the difference between  $S_{s/d}(\vec{Q})$ with ($r_s=0.5, r_n=0.5$) and without the AF+nematic fluctuations, for L=16.  (d) $\Delta S_{d}(\vec{Q}_0)$ as a function of $1/L$ for $L=14, 16 ,18$. The red dots represent the enhancement of the peak $d$-wave bond CDW structure factor  by both spin ($r_s=0.5$) and nematic ($r_n=0.5$) fluctuations . The black dots are the enhancement by spin fluctuations ($r_s=0.5$) alone. The solid curves are the best fit using $f(1/L)  = a + b/L + c/L^2$. In both cases the structure factors extrapolate to zero within error bar in the thermodynamic limit.  It suggests that the bond CDW order induced by AF and/or nematic fluctuations is short-ranged. }
\label{fig3}
\end{figure*}

\noindent{\bf{Cooper pairing induced by AF fluctuations}}\\
\noindent To investigate whether
this model with AF fluctuations [see \Eq{action} in supplementary information II] supports superconductivity we compute the equal-time pair-pair correlation function $P_{s/d}(\vec x_i)$ . Here $s/d$ denotes $s$-wave and $d$-wave pairing, respectively and $\vec x_i$ is the separation of the two pairs. In \Fig{fig2}(a) we plot the $s$- and $d$-wave pair correlation functions $\overline{P}_{d(s)}(\vec x_\textrm{max})=\frac{1}{9}\sum_{n,m=0,\pm 1} P_{d(s)}(\vec x_\textrm{max}+n\hat x + m\hat y)$ for the maximum spatial separation $\vec x_\textrm{max}=(L/2,L/2)$ for various system sizes $L$ and different values of $r_s$. (The reason for averaging over the nine neighboring values of $\vec x_\textrm{max}$ is to reduce the statistical noise).  It clearly shows that the AF spin fluctuation enhances $d$-wave pair correlation more than the $s$-wave. In particular the enhancement for $d$-wave pairing is the strongest near the AF quantum critical point $r_{s,c}\approx 0.25$.

To determine whether the ground state has SC long-range order we focus on $r_s=0.5>r_{s,c}$. (We choose this value because the relevant high $T_c$ systems do not sit at the AF quantum critical point.) We  carefully study the finite-size dependence of $\overline{P}_{d(s)}(L/2,L/2)$  for $L=8,10,12,14,16,18$. As shown in \Fig{fig2}(b) the $d$-wave pair correlation function saturates to a non-zero value after extrapolating to $L=\infty$. (The red curve is the best fit using a quadratic polynomial of $1/L$). In contrast the $s$-wave pair correlation extrapolates to zero in the thermodynamic limit. These results clearly suggest that the ground state possesses $d$-wave SC long-range order.\\

\noindent{{\bf CDW correlation induced by AF fluctuations}}\\
\noindent In the last few years resonant X-ray scattering experiments on hole doped YBCO\cite{Hayden,Keimer} and Bi2212\cite{Comin1,silva,Hashimoto} shows the existence of a short-range incommensurate CDW order. Moreover the order parameter has an approximate $d$-wave form factor\cite{Davis,Comin}, and the ordering wavevectors are consistent with those connecting nearby hot spots\cite{Comin1}. In this section we ask whether the AF fluctuation can also trigger this CDW order. Because experimentally the strongest CDW modulation is observed  among the oxygen sites (i.e. the mid points of the Cu-Cu bonds)\cite{Davis}, we focus on the bond CDW. 
We study two types of bond CDW structure factor $S_{s/d}(\vec Q)$ (see supplementary information III). Here $s/d$ denotes the $s$ and $d$-form factor. When this quantity extrapolates to a non-zero value for a particular peak wavevector $\vec{Q}_0$ as $L\to \infty$, it implies the existence of long range bond CDW order with modulation period $2\pi/|\vec{Q}_0|$. Experimentally the strongest bond CDW modulation is observed for $\vec{Q}_0=(\pm \delta,0)$ and $(0,\pm \delta)$ in the Cu-Cu bond directions with $\delta\approx 2\pi/3$\cite{Hayden,Keimer,Comin1,silva,Hashimoto}.  In \Fig{fig3}(a) we present the results of $S_{s/d}(\vec Q)$ for $L= 16$. Given the fact that the wavevectors connecting the hot spots can be both  diagonal and horizontal we scan $\vec{Q}$ in both directions. As shown in \Fig{fig3}(a) for each direction the $d$-wave bond density wave correlation is stronger than the  $s$-wave one which is consistent with the $d$-form factor found experimentally. However, disagreeing with experiments, we find the CDW ordering wavevectors lie in the diagonal directions. Our result agrees with previous approximate theoretical calculations involving hot spots\cite{Efetov,Laplaca}.

In \Fig{fig3}(b) we plot $\Delta S_{s/d}(\vec{Q})$, namely the difference of $S_{s/d}(\vec{Q})$ with and without the AF fluctuations. Here a clear peak is observed at $\vec{Q}_0=(2\pi/\lambda, 2\pi/\lambda)$ where $\lambda\approx 8a/3$. In a $L=16$ system this peak wavevector is consistent with that connecting a pair of hot spots displaced in the diagonal direction, namely $\vec{Q}_0=(2\pi/3a,2\pi/3a)$. The black dots in \Fig{fig3}(d) shows the dependence of $\Delta S_{d}(\vec{Q}_0)$ on $1/L$. The extrapolation to $L=\infty$ gives zero within errorbar hence suggesting  there is no long range CDW order. Given the fact that the ground state possesses SC long range order this should not be a surprise because these two types of symmetry breaking compete with each other.\\

\begin{figure}[t]					
\centering
\includegraphics[width=8.0cm]{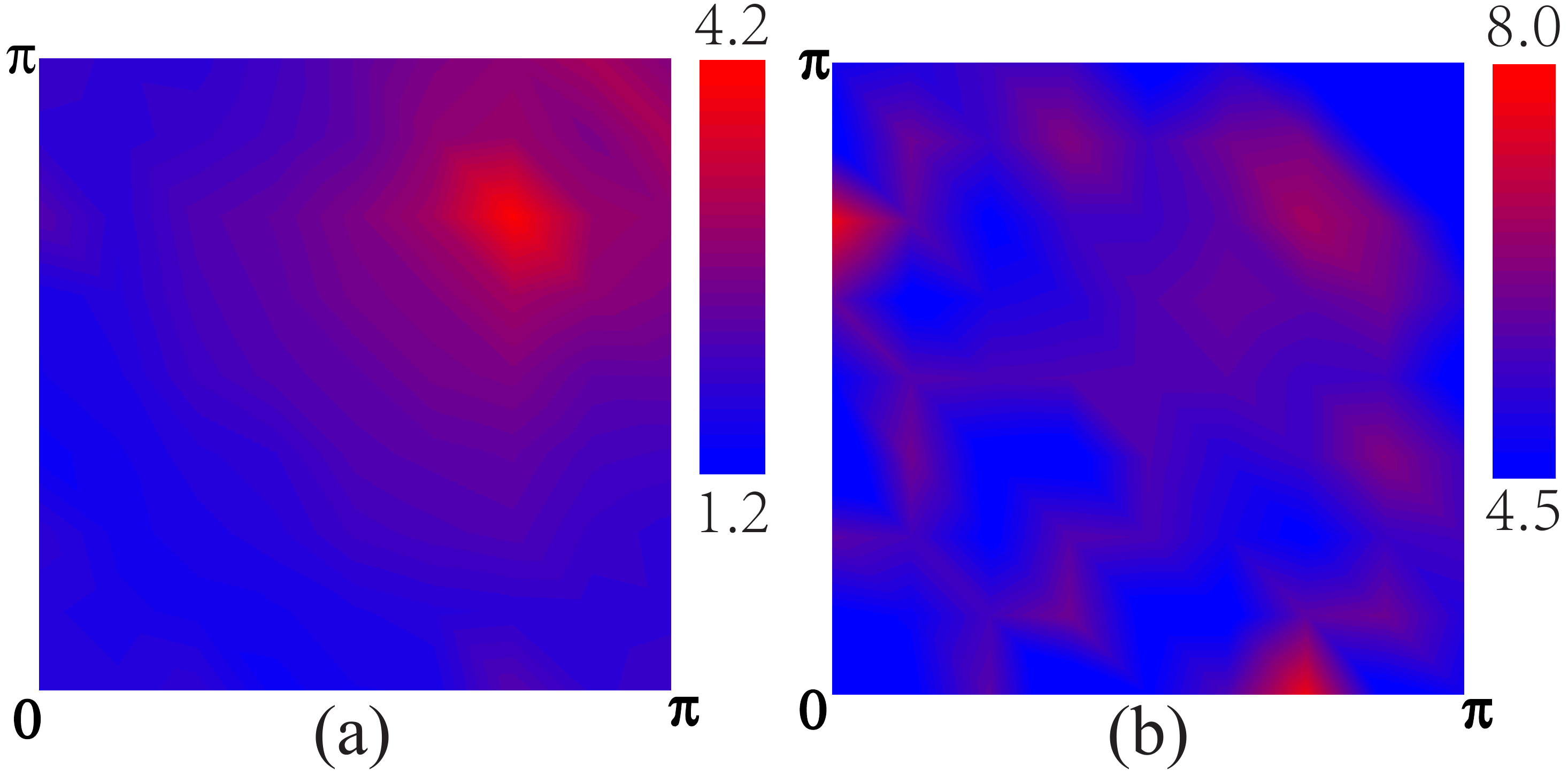}
\caption{
(a) $\Delta S_d(\vec{Q})$ for $r_s=0.5$ and $L=16$ is plotted over the first quadrant of the Brillouin zone. The peak is situated around $\vec Q=(2\pi/3,2\pi/3)$; (b) $\Delta S_d(\vec{Q})$ for $L=16$, driven by both spin ($r_s=0.5$) and nematic ($r_n=0.5$) fluctuations. The strongest peaks are situated around $\vec Q=(2\pi/3,0)$ and $(0,2\pi/3)$.  }
\label{fig4}
\end{figure}

\noindent{{\bf CDW correlation induced by both AF and nematic fluctuations}}\\
\noindent The fact the predicted directions of the CDW ordering wavevector are inconsistent with experiments makes us to suspect
that AF fluctuations alone is insufficient to account for the intertwined orders in the cuprates.
Motivated by the fact that nematicity have been observed  in many cuprates\cite{Ando,Hinkov,Lawler,Daou}, we add the nematic fluctuations
to the effective action (see supplementary information IV). The parameter that controls the strength of the nematic fluctuations is $r_n$. Large $r_n$ causes the nematic order parameter to become disordered.
In \Fig{fig3}(c) we plot $S_{s/d}(\vec{Q})$ after the inclusion of the nematic fluctuations. Here we choose $r_s=0.5,r_n=0.5$ which is on the disorder side of both the AF and nematic phase transitions. The results show the $d$-wave bond CDW with wavevectors $\vec{Q}=(\pm 2\pi/\lambda,0)$ and $(0,\pm 2\pi/\lambda)$ where $\lambda\approx 8a/3$  becomes  the dominant  instability ! The $S_d(\vec{Q})$  without and with the nematic fluctuations are plotted over the entire Brillouin zone in \Fig{fig4}(a) and \Fig{fig4}(b). These plots confirm the global maximum of $\Delta S_d(\vec{Q})$ in the whole Brillouin zone indeed locates at the previously described locations.

\begin{figure}[t]						
\centering
\includegraphics[width=8.0cm]{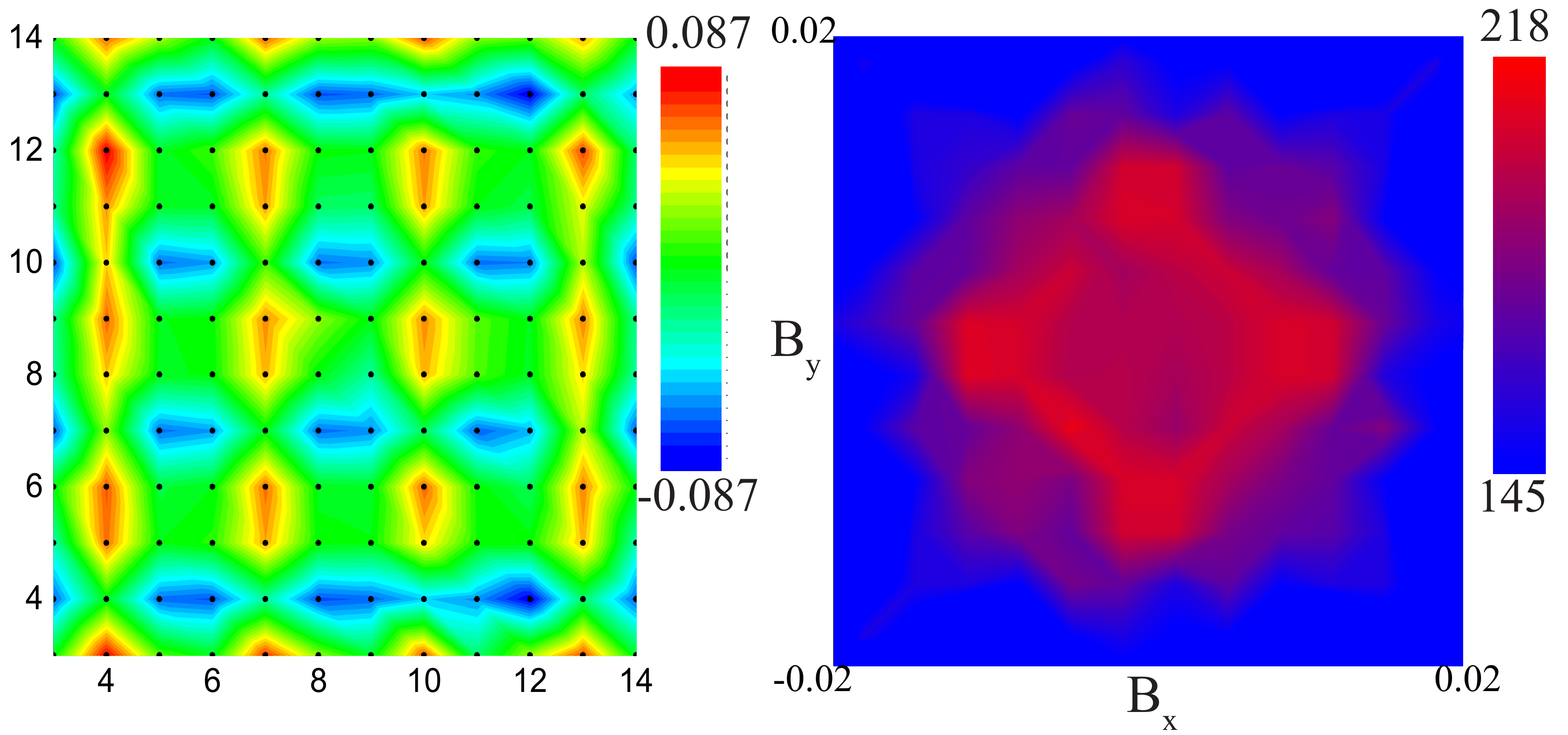}
\caption{
(a) The $d$-wave bond CDW order parameter $\langle B_d(i)\rangle \equiv \langle \sum_{a=\pm \hat x}\psi^{\dagger}_{i,x}\s_0\psi_{i+a,x}- \sum_{a=\pm hat y}\psi^{\dagger}_{i,y}\s_0\psi_{i+a,y}\rangle$ in a system of size $L=16$ with open boundary conditions. Here $\s_0$ is the identity matrix in the spin space. The black dots represent the lattice sites. The pattern of bond CDW order parameter modulations obtained numerically is almost perfectly consistent with the expected modulations with 3a period.
(b) The concurrence probability $P(B_x,B_y)$ in the QMC simulations is plotted as a function of  $B_x$ and $B_y$, where $B_{x/y}$ are the $d$-wave bond CDW order parameters associated with ordering wavevector $(2\pi/3,0)$ and $(0,2\pi/3)$, respectively. The system size in this computation is $L= 15$, and the parameters $(r_s,r_n)$ used is $(0.5,0.5)$.}
\label{fig6}
\end{figure}
Even with the nematic fluctuations the CDW correlation is short-range. This is shown by the red points of \Fig{fig3}(d) which plots $\Delta S_d(\vec{Q}_0)$ as a function of $1/L$, again the extrapolation to the thermodynamic limit suggests there is no CDW long range order. However even short-range CDW correlations can imply a substantial CDW susceptibility. In its presence an actual CDW pattern may be induced when there is external translation symmetry breaking perturbations such as quenched disorders. Such kind of patterns can be observed in an STM experiment. In \Fig{fig6}(a),  we show a static CDW pattern induced by the open boundary condition in the system of size $L=16$. The action used to generate this pattern has sufficiently strong nematic fluctuations so that the CDW ordering wavevectors are in the horizontal and vertical directions.  The CDW modulation wavevector is about $2\pi/3a$ agreeing with the momentum distance between two horizontally displaced hot spots. In addition the $d$-wave nature of the bond CDW is transparent.

An open and interesting issue concerning the CDW is whether it is uni- or bi-directional. The answer relies on the sign of the quartic coupling between the horizontal and vertical CDW order parameters in the Ginzburg-Landau action. Since the quartic term only becomes significant when the magnitude of the order parameter is appreciable, it is difficult to answer this question by watching the induced CDW pattern, such as that in \Fig{fig6}(a), in a system without the CDW long range order. However in a Monte-Carlo simulation we can
determine the concurrence probability
 $P(B_x,B_y)$ where $B_{x,y}$ are the CDW order parameters associated with $\vec{Q}=(\pm 2\pi/3a,0)$ and $(0, \pm 2\pi/3a)$, respectively. If the quartic coupling favors the uni-directional CDW, the peaks of $P$ should appear on the horizontal and vertical axes of the $(B_x,B_y)$ plane. Conversely if the quartic coupling favors the bi-directional CDW the peaks should appear along the  diagonal direction. In \Fig{fig6}(b) we plot $P$ over the $(B_x,B_y)$ plane. Four peaks on horizontal and vertical axes are seen, implying the quartic coupling term favors the uni-directional (stripe) CDW.
\\

\noindent{{\bf The effects of nematic fluctuation on pairing}}\\
\noindent Finally we return to superconductivity. Specifically we study the effects of nematic fluctuation on the SC order.  In \Fig{fig2}(c), we plot the SC pair correlation at $\vec x_\textrm{max}$ for L=14, with and without nematic order parameter fluctuations, as a function of $r_s$. The $d$-wave SC pairing correlations is moderately enhanced in the presence of nematic fluctuations $(r_n=0.5)$. In \Fig{fig2}(d), we perform a finite-size scaling of the $s$- and $d$-wave pair correlation functions with both AF and nematic fluctuations ($r_s=r_n=0.5$). The extrapolated $d$-wave SC order paramater is greater than the value induced by the AF fluctuation alone. We have also studied SC in the nematic long range ordered phase. Interestingly even under that condition the SC long range order is significantly enhanced. Because in the nematic phase the $d$ and $s$ wave SC order parameters can mix, in \Fig{S1}(a) and \Fig{S1}(b) of  supplementary information V we show the $L$-dependence of both order parameters. When extrapolated to the thermaldynamics limit the $d$-wave SC order parameter is much stronger than the $s$-wave one.

To conclude, our intrinsically unbiased QMC study of an effective theory involving the same hot spots as the cuprates clearly indicates that in order to describe the intertwined orders in the cuprates nematic fluctuations are indispensable. Remarkably, the coupling to nematic fluctuations not only gives rise to correct $d$-form factor bond CDW ordering but also  enhances the $d$-wave superconducting long-range order. Assuming that hot spot based effective theories could capture the essential low-energy physics of the cuprates, our results significantly further the understanding of pairing mechanism in cuprates.\\

{\noindent\bf Acknowledgement} \\
\noindent We would like to thank Seamus Davis and Steve Kivelson for helpful discussions. ZXL and HY were supported in part by the National Thousand Young-Talents
Program and the NSFC under Grant No. 11474175.
FW was supported by the National Science Foundation of China(Grant No. 11374018).
DHL was  supported  by  the  U.S.  Department  of  Energy,  Office  of  Science,  Basic
Energy Sciences, Materials Sciences and Engineering Division, grant DE-AC02-05CH11231.

\begin{widetext}
\section{Supplementary Information}

\renewcommand{\theequation}{S\arabic{equation}}
\setcounter{equation}{0}
\renewcommand{\thefigure}{S\arabic{figure}}
\setcounter{figure}{0}
\renewcommand{\thetable}{S\arabic{table}}
\setcounter{table}{0}

\subsection{I. The AF effective action}
The AF effective action based on the two-band model is given by $S=S_F+S_s$ where
\bea\label{action}
S_F &=&\int_0^\beta d\tau\Big\{ \sum_{ij,\alpha=x,y} \psi_{i\alpha}^{\dagger}\big[(\partial_\tau -\mu)\delta_{ij}-t_{ij,\alpha}\big]\psi_{j\alpha} +\lambda_s\sum_i (-1)^i \Big[\psi^{\dagger}_{ix}(\vec{\sigma}\cdot\vec{\varphi}_i)\psi_{iy}+h.c.\Big]\Big\}
   ,\\
S_{s} &=& \int_0^\beta d\tau\Big\{ \frac{1}{2}\sum_i \frac{1}{c_s^2}|\partial_\tau \vec{\varphi_i}|^2 + \frac{1}{2}\sum_{\langle ij\rangle}|\vec{\varphi_i}-\vec{\varphi}{_j}|^2
  +\sum_i\Big[\frac{r_s}{2}|\vec{\varphi}_i|^2+\frac{u_s}{4}\left(|\vec{\varphi}_i|^2\right)^2\Big]\Big\}.
\label{action2}
\eea
Here $i$ labels the sites of a square lattice, $\alpha=x,y$ labels the two orbitals (which transform into each other under the 90$^\circ$ rotation) from which the red and blue Fermi surfaces in \Fig{fig2}(b) are derived from, $\tau$ denotes the imaginary time and $\beta$ is the inverse temperature. In \Eq{action} $\vec{\varphi}$ is the Neel order parameter and the operator $\psi_{i\alpha}$ is a spinor operator which annihilates an electron in  orbital $\alpha$ and on site $i$. The three $\vec{\sigma}$ are the spin Pauli matrices.

The parameters in this effective action include the spin wave velocity $c_s$, and $r_s$ which tunes the system across the AF phase transition, and $u_s$ is the self-interactions of the $\vec \phi$ field, and $\lambda_s$ is the ``Yukawa'' coupling between the electrons and the AF fluctuation. The hopping integral $t_{ij}$ is chosen to be among nearest neighbor sites and equal to $t_{\parallel}= 1.0$ for $x(y)$-orbital along $x(y)$ direction and $t_\perp = 0.5$ for $y(x)$ orbital along $x(y)$ direction. We choose the chemical potential to be $\mu=-0.5$ such that the Fermi surface is shown in \Fig{fig1}(b).

Aside from the square lattice spatial symmetries the action is invariant under the anti-unitary transformation $U = i\tau_z\sigma_y K$, where $\tau_z$ is the third Pauli matrix acting in orbital space and $K$ denotes complex conjugation. It can be shown that because of this symmetry, the fermion determinant for arbitrary $\vec{\varphi}_i(\tau)$ configuration is positive hence the QMC simulation is free of minus-sign. This enables us to perform large-scale projective QMC simulation. In the calculation we fixes $\lambda_s$, $c$ and  $u_s$ to unity, and vary the value of $r_s$ to control the severity of AF fluctuation. From the computed Binder cumulant associated with the AF order parameter we determined the AF quantum critical point to situate at $r_{s,c}=0.25\pm 0.1$, consistent with the results of Berg {\it et al}.

\subsection{II. The superconducting pair correlation function}

To investigate whether Eq.(S1) and Eq.(S2) support superconductivity we compute the equal time pair-pair correlation functions \bea P_{s/d}(\vec r_i)= \langle\Delta_{s/d}(\vec r_i)\Delta_{s/d}^{\dagger}(\vec 0)\rangle\eea  where \bea\Delta_{s/d}(\vec r_i)= \psi^T_{ix}(i\s_y)\psi_{ix}\pm\psi^T_{iy}(i\s_y)\psi_{iy}\eea are the $s$ (+ sign) and $d$ ($-$ sign) wave Cooper pair operators, respectively.

\subsection{III. The bond CDW structure factor}

We define the bond CDW operator at site $i$ as
\bea B_{s/d}(i)=\sum_{a=\pm \hat x}\big[\psi^{\dagger}_{ix}\s_0 \psi_{i+a,x}+H.c.\big]\pm \sum_{a=\pm\hat y}\big[\psi^{\dagger}_{iy}\s_0\psi_{i+a,y}+H.c.\big],
\eea
where $\s_0$ is the identity $2\times 2$ matrix and $+$ sign corresponds to the $s$-wave and $-$ to the $d$-wave form-factor, respectively.
Under a 90$^\circ$ rotation around site $i$, $B_s(i)\ra B_s(i)$ while $B_d(i)\ra -B_d(i)$. The bond CDW structure factor is defined as  \bea S_{s/d}(\vec Q)=\frac{1}{N^2}\sum_{ij}\langle B_{s/d}(i) B_{s/d}(j)\rangle\cos[\vec Q\cdot (\vec x_i-\vec x_j)].
\eea

\subsection{IV. Adding the nematic fluctuation to the effective theory}

Adding nematic fluctuation to the effective theory amounts to
$S\ra S+\Delta S$ where
\bea
&&\Delta S=\lambda_n\int_0^\beta d\tau\sum_i \chi_i \Big[\psi^{\dagger}_{ix}\s_0\psi_{ix}-\psi^{\dagger}_{iy}\s_0\psi_{iy}\Big]+S_n\nonumber \\
&&S_n=\int_0^\beta d\tau\Big\{ \frac{1}{2}\sum_i \frac{1}{c_n^2}|\partial_\tau \chi_i|^2 + \frac{1}{2}\sum_{\langle ij\rangle}|\chi_i-\chi_j|^2 +\sum_i\Big[\frac{r_n}{2}|\chi_i|^2+\frac{u_n}{4}\chi_i^4\Big]\Big\}.
\eea
In the simulation we choose $r_n=0.5(0.0)$ on the disorder (ordered) side of the nematic phase transition and $c_n=u_n=\lambda_n=1$.

\subsection{V. Superconductivity in the nematic ordered phase}
Due to the likelihood that the superconductivity in the hole doped cuprates occurs in the nematic ordered phase, in this section we study the coexistence of superconductivity and nematic long-range order. The results are shown in \Fig{S1}. It is worth to mention that in the nematic ordered phase the $d$ and $s$ wave pairing can mix. This is shown in \Fig{S1}(b) where a minor $s$-wave component coexists with the dominant $d$-wave component. Such mixture has been observed in YBCO where the four-fold rotation symmetry is explicitly broken by the copper-oxygen chains\cite{clarke}.
\begin{figure*}[h]
\includegraphics[width=10.0cm]{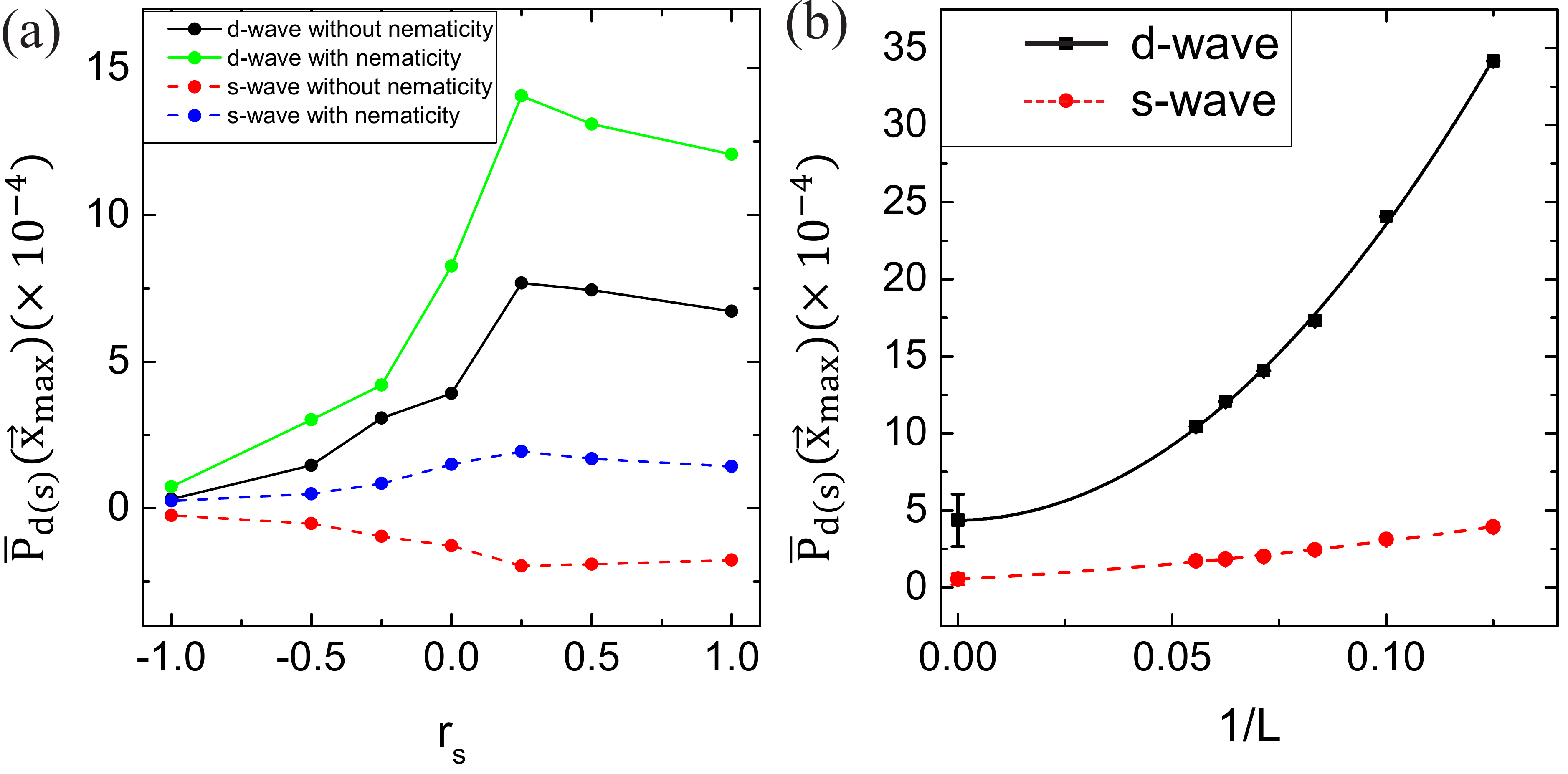}
\caption{
(a) The comparison of the $d$- and $s$-wave SC pairing correlations in the system of $L=14$, without and with nematic long range order. In the latter case, we set $r_n=0$. These results show that the $d$-wave SC pairing is enhanced even in the nematic long-range ordered phase. (b) We plot $\overline P_{d(s)}(\vec x_\textrm{max})$ in the nematic long-range ordered phase versus $1/L$ for $L=8,\cdots,18$. The solid and dashed lines are the best fit using $f(1/L) = a + b/L + c/L^2$. Besides the dominant $d$-wave pairing long-range order, a weak but finite $s$-wave component also emerges, as expected in the nematic phase.  }
\label{S1}
\end{figure*}

\subsection{VI. Projector Quantum Monte Carlo}
Projector quantum Monte Carlo\cite{Sorella,White} is one of the determinant QMC algorithms\cite{Scalapino} to investigate the ground state properties of a quantum many-body model. In projector QMC, the expectation value of an observable in the  ground state can be evaluated as:
\bea
\frac{ \bra{\psi_{0}} O \ket{\psi_0}}{ \avg{\psi_{0} \mid \psi_{0} } }   = \lim_{\theta\rightarrow \infty} \frac{ \bra{\psi_T}e^{-\theta H } O e^{-\theta H} \ket{\psi_T}}{ \bra{\psi_T} e^{-2\theta H} \ket{\psi_T}}
\eea
where $ \psi_0 $ is the true ground state wave function and $ \psi_T$ is a trial wave function which we assume has a finite overlap with ground state wave function. The imaginary-time projection parameters in our computation is $\Theta = 40/t$, which is sufficient to obtain converged ground-state quantities within statistical uncertainty. We set the imaginary step $\Delta\tau = 0.1$ and also check that results do not change using smaller $\Delta\tau$. Here, $Z_T \equiv   \bra{\psi_T} e^{-2\theta H} \ket{\psi_T}$ plays the role of partition function.  After discretizing imaginary time with standard Trotter-Suzuki decomposition\cite{Assaad}, it can be expressed in the form of path integral :
\bea
Z_T = \int  [D \vec{\varphi}]
[D \chi] e^{ - S_B[ \vec{\varphi},\chi]}
\bra{\psi_T} \prod_{\tau=1}^{N}\hat{B_\tau} \ket{\psi_T}, \nonumber \\
\nonumber\\
\eea
where
\bea
S_B[ \vec{\varphi},\chi]&=&\int_0^\beta d\tau\Big\{ \frac{1}{2}\sum_i \frac{1}{c_s^2}|\partial_\tau \vec{\varphi}_{i}|^2 + \frac{1}{2}\sum_{\langle ij\rangle}|\vec{\varphi}_{i}-\vec{\varphi}_{j}|^2
+\sum_i\Big[\frac{r_s}{2}|\vec{\varphi}_{i}|^2+\frac{u_s}{4}\left(|\vec{\varphi}_{i}|^2\right)^2\Big]\Big\}\nonumber \\&+&\int_0^\beta d\tau\Big\{ \frac{1}{2}\sum_i \frac{1}{c_n^2}|\partial_\tau \chi_i|^2 + \frac{1}{2}\sum_{\langle ij\rangle}|\chi_i-\chi_j|^2 +\sum_i\Big[\frac{r_n}{2}|\chi_i|^2+\frac{u_n}{4}\chi_{i}^4\Big]\Big\}.
\label{Bose}
\eea
In \Eq{Bose} $\vec{\varphi}$ and $\chi$ are the Bose fields associated with the antiferromagnetic and nematic fluctuations, respectively.

The operator $\hat{B_\tau}$ are given by:
\bea
\hat{B_\tau} = e^{-\frac{1}{2}\Delta\tau \psi^\dagger K \psi} e^{-\Delta\tau \psi^\dagger V^s_\tau \psi}
e^{-\Delta\tau \psi^\dagger V_\tau^n\psi} \nonumber \\
\nonumber \\
\eea
Matrices $K$,$V^s_\tau$ and $V_\tau^n$  are:
\bea
&&K_{\tau;ij;\alpha,\alpha';s,s'} = (\tau_0)_{\alpha,\alpha'} (\sigma_0)_{s,s'} (-t_{\alpha,ij}-\mu) \nonumber \\
&&V_{\tau;ij;\alpha,\alpha';s,s'}^s = \lambda_s
(\tau_1)_{\alpha,\alpha'}\delta_{ij}[(\vec{\sigma})_{s,s'}\cdot\vec{\varphi}_{i}(\tau)] \nonumber \\
&&V_{\tau;ij;\alpha,\alpha';s,s'}^n = \lambda_n
(\tau_3)_{\alpha,\alpha'}(\sigma_0)_{s,s'}\delta_{ij}\chi_i(\tau).
\eea
Here, $\tau_{0,1,2,3}$ are the Pauli matrices in orbital ($x,y$) space and $\sigma_{0,1,2,3}$ are the Pauli matrices in spin space. In addition $\alpha,\alpha' = x,y$ are orbital indices, and $s,s' = \uparrow,\downarrow$ are spin indices. We choose the fermion trial wave function as:
\bea
\ket{\psi_T} = \prod_{a=1}^{N_f} ( c^{\dagger} P )_a \ket{0},
\eea
where P is a matrix with $ N\times N_f$ dimension. $N$ is the number of sites and $N_f$ is the number of fermions. Usually, the trial wave is generated from the ground state of non-interacting Hamiltonian. We always check the choice of trial wave function has no influence on results if $\theta$ is large enough.

After fixing trial wave function, the expectation value $\bra{\psi_T} \prod_{\tau=1}^{N}\hat{B_\tau} \ket{\psi_T}$ can be evaluated\cite{White}:
\bea
\bra{\psi_T} \prod_{\tau=1}^{N}\hat{B_\tau} \ket{\psi_T} =  \det[ P^{\dagger} \prod_{\tau=1}^{N} B_\tau P],
\eea
thus the effective quantum partition function in projector QMC can be expressed as:
\bea
Z_T = \int [D \vec{\varphi}]
[D \chi] e^{ - S_B[ \vec{\varphi},\chi]}
\det[ P^{\dagger}\prod_{\tau=1}^{N} B_\tau P]
\eea
In our simulation different configurations of the $\vec{\varphi}_{i}(\tau)$,
 $\chi_{i}(\tau)$ fields are sampled using standard Monte Carlo techniques.

\end{widetext}

\end{document}